\documentstyle[osa,manuscript,psfig]{revtex} 

\begin{document}                                                          
\titlepage

\title{ 
Self-organized criticality and interacting soft gluons in deep-inelastic
electron-proton scattering}
\author { C. Boros\footnote{
             Now at Centre for Subatomic Structure of Matter (CSSM),
          University of Adelaide, Australia 5005.}, 
          Meng Ta-chung, R. Rittel and Zhang Yang\\
 {\it Institut f\"ur Theoretische Physik, FU Berlin,}\\
 {\it 14195 Berlin, Germany}} 
 \maketitle
                                                       
\begin{abstract}
\noindent
It is suggested that the colorless systems of
interacting soft-gluons in large-rapidity-gap 
events are open dynamical complex systems
in which self-organized criticality and BTW-clusters
play an important role. Theoretical arguments and 
experimental evidences supporting such a statistical
approach to deep-inelastic scattering are presented.
\end{abstract}

\newpage

\subsection*{I. Interacting soft gluons in the small-$x_B$
                region of DIS}
A number of striking phenomena have
been observed in recent 
deep-inelastic electron-proton scattering experiments  
in the small-$x_B$ region. In particular it is seen, 
that the contribution of the gluons dominates\cite{r1},
and that large-rapidity-gap (LRG) events exist\cite{r2}. 
The latter shows that the virtual photons in such processes may
encounter colorless objects originating from the proton.

The existence of LRG events in these scattering processes
have attracted much attention, and
there has been much discussion\cite{r2,r3,r4,r5,r6,r7,r8}
on problems associated with the origin and/or the 
properties of such colorless objects.
Reactions in which ``exchange'' of such objects dominate
are known in the literature\cite{r3,r4,r5} as 
``diffractive scattering processes''.
While the concepts and methods used by different 
authors  are in general very much different from one another,
all the authors in describing such processes
(experimentalists as well as theorists)
seem to agree on the following\cite{r5}
(see also Refs.
[\ref{r2}--\ref{r4}, \ref{r6}--\ref{r8}]):
(a) Interacting soft gluons play a dominating role in
understanding the phenomena in the small-$x_B$ region in 
general, and in describing the properties of LRG events in particular.
(b) Perturbative QCD should be, and can be, used to describe the 
LRG events associated with high transverse-momentum ($p_\perp$)
jets which have been observed at HERA\cite{r6} and at the Tevatron\cite{r7}. 
Such events are, however, rather rare.
For the description of the bulk of LRG events, concepts 
and methods beyond the perturbative QCD (for example,
Pomeron Models\cite{r4} based on Regge Phenomenology) are needed.
The question, whether or how
perturbative QCD plays a role in such non-perturbative approaches does
not have an unique answer.

In a previous paper\cite{r8}, we suggested that the observed dominance of 
interacting soft gluons\cite{r1} and the existence of LRG events\cite{r2}
in the small-$x_B$
region are closely  related to each other, and that the interacting soft 
gluons may form colored 
and colorless systems --- which we called 
``gluon clusters''. Such gluon clusters have finite lifetimes 
which (in the small-$x_B$ region) can be of the same order as the 
interaction time $\tau_{\rm int}$ --- the time-interval in which the 
virtual photon $\gamma^\star$ ``sees'' 
the cluster in the sense that it is absorbed by
the charged 
constituents of
the latter.
In Ref.[\ref{r8}]
the lifetime of such a 
gluon-cluster was {\em estimated} by using uncertainty principle and
kinematical considerations --- without any {\em dynamical} input.
In analogy to hadron structure function, a quantity $F_2^c$ which we called
``the structure function of the gluon cluster $c_0^\star$'' was introduced,
and then it was set to be a constant --- in accordance with
the purpose of that paper which is to discuss the {\em kinematical} 
aspects of a statistical approach.

After having seen what phase-space considerations can, and cannot do,
we decided to go one step further, and study {\em the dynamical aspects}
of the interacting soft-gluons in these scattering processes. 
In doing so,
we realized that the system of interacting 
soft-gluons is
extremely complicated. It is not only too complicated (at least for us) 
to take
the details of local interactions into account 
(for example by describing
the reaction mechanisms in terms of Feynman diagrams),
but also too complicated to apply well-known concepts
and methods in conventional equilibrium statistical mechanics.
In fact, having the above-mentioned empirical facts about LRG events
and the basic properties of gluons prescribed by the QCD-Lagrangian
in mind, we are readily led to the following picture:

Such a system is
{\it an open  dynamical system with many degrees of freedom},
and it is in general {\em far from equilibrium}.
This is because, once we accept that the colorless object (which the 
virtual photon encounters) is a system of soft gluons whose interactions
are not negligible, we are also forced to accept
that, in such a system,
gluons can be emitted and absorbed 
by the members of the system as well as
by gluons and/or  quarks and antiquarks outside the system
(we note in particular that, since the gluons are soft,
their density in space is high,
and the distances between the interacting gluons are in general
not short, the ``running-coupling-constant''
can be very large). Furthermore, since in general more than one
gluons can be emitted or absorbed
by the members of the system, the system itself can remain to be a 
color-singlet. This means in particular that, in such a system, 
{\it neither the
number of gluons nor the energy of the system can be a 
conserved quantity}. 

Do we see comparable open, dynamical, complex systems in Nature?
If yes, what
are the characteristic 
features of such systems?

\subsection*{II. Characteristic features of open dynamical complex systems}
Open dynamical complex systems are not difficult to find in Nature ---
at least not in the macroscopic world! Such systems have been studied,
and in particular the following have been observed by
Bak, Tang and Wiesenfeld (BTW) some time ago\cite{r9}:
Open 
dynamical systems with many degrees of freedom may
evolve to 
self-organized critical states which lead to
fluctuations extending over all length- and
time-scales, and that 
such fluctuations manifest themselves in form of
spatial and temporal power-law scaling behaviors 
showing properties
associated with fractal
structure and flicker  noise respectively. 

BTW\cite{r9} and many other
authors\cite{r10} proposed, and demonstrated by
numerical simulations, the following:  Dynamical systems with local
interacting degrees of freedom can evolve into self-organized
structures of states which are barely stable. A local perturbation of a
critical state may ``propagate'', in the sense that it spreads to (some)
nearest neighbors, and than to the next-nearest neighbors, and so on in
a ``domino effect'' over all length scales,  the size of 
such an ``avalanche'' can be as
large as the  entire
system. Such a ``domino effect'' eventually terminates after a total time $T$,
having reached a final amount of dissipative energy and having
effected a total spatial extension $S$. The quantity $S$ is called by
BTW the ``size'', and the quantity $T$ the ``lifetime'' of the 
avalanche --- named by BTW a ``cluster''
 (hereafter referred to as BTW-cluster). As  we
shall see in more details later on, it is of considerable importance to
note that a BTW-cluster {\it cannot}, and {\it  should not}
be identified with a cluster in the usual sense. 
It is an avalanche, 
{\it not} a {\it static} object 
with a fixed structure
which remains unchanged until it
decays after a time-interval (known as the lifetime in
the usual sense). 

It has been
shown\cite{r9,r10} that the 
distribution ($D_S$) of 
the ``size'' (which is a measure of
the dissipative energy, $S$) and the distribution 
($D_T$) of the lifetime
($T$) of BTW-clusters in  such open 
dynamical systems obey   power-laws:
\begin{equation}
\label{e1}
D_S(S)\sim S^{-\mu},
\end{equation}
\begin{equation}
\label{e2}
D_T(T)\sim T^{-\nu},
\end{equation}
where $\mu$ and $\nu$ are positive real constants. In fact, such spatial and 
temporal power-law
scaling behaviors can be, and have been, considered
as the universal signals --- the ``fingerprints'' ---   of  the 
locally  perturbed
self-organized  critical states in such systems. 
It is
expected\cite{r9,r10} that the general concept of self-organized
criticality (SOC), which  is 
complementary to chaos, may be 
{\it the} underlying concept for temporal and spatial scaling in a wide class
of {\it open non-equilibrium systems} --- although it is not yet known 
how the exponents in such power law can be calculated analytically.

SOC has been observed in 
a large number of open dynamical complex systems in
non-equilibrium\cite{r9,r10,r12,r13,r14,r15} 
among which the following examples are
of particular interest, because they illuminate several aspects of
SOC which are relevant for the discussion in this paper.

First, the well known Gutenberg-Richter law\cite{r11,r12}
for earthquakes as a special
case of Eq.(1): 
In this case, $S$ stands for the released energy (the magnitude)
of the earthquakes. $D_S(S)$ is the number of
earthquakes at which an energy $S$ is released.
Such a simple law is known to be valid
for all earthquakes, large (up to $8$ or $9$ in Richter scale)
or small! We note, the power-law behavior given by the
Gutenberg-Richter law implies in particular the following.
The question ``How large is a typical earthquake?'' does
not make sense!

Second, the sandpile experiments\cite{r9,r10} which show 
the simple regularities mentioned in Eqs.(1) and (2): 
In this example, we see how local perturbation can be caused by the 
addition of one grain of sand (note that we are dealing with 
an open system!). Here, 
we can also see how
the 
propagation of perturbation in form of ``domino effect'' 
takes place, and 
develops into avalanches of all possible sizes and durations.
The size- and duration-distributions are given by Eqs.(1) 
and (2) respectively.
This example is indeed a very attractive one,
not only because such 
experiments can be, and have been performed in labs\cite{r10}, but also 
because they can
be readily simulated on a PC\cite{r9,r10}.

Furthermore, it has been pointed out, and demonstrated
by simple models\cite{r10,r13,r14,r15},
that the concept of SOC can also be applied
to Biological
Sciences.
It is amazing to see how phenomena as complicated as Life 
and Evolution can be simulated
by simple models such as the ``Game of Life''\cite{r13} and
the ``Evolution Model''\cite{r14,r15}.

Having seen that systems of interacting soft-gluons
are open dynamical complex systems,
and that a wide class of open systems with many degrees of 
freedom in the macroscopic world
evolve to self-organized critical states which lead to
fluctuations extending over all length- and time-scales,
it seems natural to ask the following:
Can such states and such fluctuations
also exist in the microscopic world --- on the
level of quarks and gluons?

\subsection*{III. Are  gluon-clusters hadron-like?} 
How can we find out whether the general concept 
of self-organized criticality
(mentioned in Section II)
plays a role
in diffractive deep-inelastic lepton-hadron scattering
processes (discussed in Section I)? 
A simple and effective way 
of doing this, is to check whether the ``fingerprints''  
mentioned in Eqs.(~\ref{e1}) and (~\ref{e2}),
which can be considered as the necessary conditions for
the existence of self-organized criticality, show up 
in the relevant 
experiments. 
For such a comparison, we need  
the spatial and the temporal distributions of the gluon-clusters. 
Hence, an important step in our quantitative study is
to obtain these distributions directly from
the experimental data --- if possible, without 
{\em any} theoretical input.
Having this goal in mind, we now try to express such 
cluster-distributions in terms of
the measured \cite{r3}
``diffractive structure function'' 
\mbox{$F_2^{D(3)}(\beta,Q^2;x_P)\equiv \int dt F_2^{D(4)}(\beta,Q^2;x_P,t)$}.
Here, we note that $F_2^{D(4)}(\beta,Q^2;x_P,t)$ 
is related \cite{r3,r4,r5,r6} to the 
differential cross-section for large-rapidity-gap
events 
\begin{equation}
\label{a3}
{d^4\sigma^D\over d\beta dQ^2 dx_P dt}={4\pi\alpha^2\over\beta
Q^4}(1-y+{y^2\over 2})F_2^{D(4)}(\beta,Q^2;x_P,t),
\end{equation}
in analogy to 
the relationship between the corresponding quantities
[namely $d^2\sigma/(dx_B\,dQ^2)$ and $F_2(x_B,Q^2)$]
for normal deep-inelastic electron-proton scattering events
\begin{equation}
\label{a4}
{d^2\sigma\over dx_BdQ^2}={4\pi\alpha^2\over
x_BQ^4}(1-y+{y^2\over 2})F_2(x_B,Q^2).
\end{equation}
The kinematical variables, in particular $\beta$, $Q^2$, $x_P$ and $x_B$ 
(in both cases) are directly measurable quantities, the  definitions 
of which are shown in Fig.1 together with the corresponding
diagrams of the
scattering processes. We note
that, although these variables are 
Lorentz-invariants, it is sometimes convenient to interpret them in a
``fast moving frame'', for example the electron-proton center-of-mass 
frame  where the proton's 3-momentum $\vec P$ is large (i.e. its 
magnitude $|\vec P|$ and thus the energy $P^0\equiv (|\vec P|^2+M^2)^{1/2}$ 
is much larger than the proton mass $M$). While $Q^2$ characterizes 
the virtuality of the space-like photon
$\gamma^\star$, $x_B$ can be interpreted, 
in  such a ``fast moving frame'' (in the framework 
of the celebrated parton model),  as the
fraction of proton's energy $P^0$ (or longitudinal momentum $|\vec P|$)
carried by the struck charged constituent. 

We recall, in the framework 
of the  parton model, $F_2(x_B, Q^2)/x_B$ for ``normal events''
can be interpreted  as the sum of the probability densities 
for the above-mentioned $\gamma^\star$ to interact with such 
a  charged constituent inside the proton. In analogy to this, 
the quantity
$F_2^{D(3)}(\beta,Q^2;x_P)/\beta$ for LRG events
can be interpreted  as the sum of the probability 
densities for $\gamma^\star$ to interact with 
a charged constituent which 
carries a fraction $\beta\equiv x_B/x_P$ of the energy (or longitudinal 
momentum) of  the colorless object,
under the  condition that the colorless object
(which we associate with a system of interacting soft gluons) carries a
fraction $x_P$
of proton's energy (or longitudinal momentum). 
We hereafter denote this
charged-neutral and color-neutral gluon-system by
$c^\star_0$ (in Regge pole models\cite{r4} this object  is known as 
the ``pomeron''). 
Hence, by comparing Eq.\,(3) with Eq.\,(4) and by comparing the two 
diagrams shown in Fig.\,1(a) and Fig.\,1(b), it is tempting to draw
the following conclusions: 

The diffractive process is nothing else but
a process in which the virtual photon $\gamma^\star$ 
encounters a $c_0^\star$,
and $\beta$ is nothing else but the Bjorken-variable with respect to 
$c_0^\star$ (this is why it is called $x_{BC}$ in Ref.[\ref{r8}]). 
This means, 
a diffractive $e^-p$ scattering event can be envisaged as an event in 
which the virtual photon $\gamma^\star$ collides with ``a $c_0^\star$-target'' 
instead of ``the  proton-target''. 
Furthermore, since  $c_0^\star$ is charge-neutral,
and  a  photon can only directly interact with an object
which has electric charges and/or magnetic moments,
it is tempting to assign $c_0^\star$ an
electromagnetic structure function $F_2^{c}(\beta, Q^2)$,
and study the interactions between the virtual photon and the quark(s)
and antiquark(s) inside $c_0^\star$.
In such a picture
(which should be formally the same as that of
Regge pole models\cite{r4},
if we would replace the $c_0^\star$'s by ``pomerons'')
we are confronted with the following two questions: 

First, is it possible and meaningful to discuss the $x_P$-distributions of 
the $c_0^\star$'s without knowing the intrinsic properties, in particular the 
electromagnetic structures, of such objects? 

Second,are gluon-clusters hadron-like, such that their electromagnetic 
structures can be studied 
in the same way as those for
ordinary hadrons?

We discuss the second question here, and leave the first question to
the next section. 
We note, in order to be able to answer the second question
in the {\em affirmative}, 
we need to know {\em whether}
$F_2^{D(3)}(\beta,Q^2;x_P)$ can be factorized in the form
\begin{equation}
\label{eee1}
F_2^{D(3)}(\beta, Q^2;x_P)=f_c(x_P)F_2^c(\beta,Q^2).
\end{equation}
Here, $f_c(x_P)$ plays the role of a ``kinematical factor''
associated with the ``target $c_0^\star$'', 
and $x_P$ is  the fraction
of proton's energy (or longitudinal momentum) carried by
$c_0^\star$. [We could call $f_c(x_P)$
``the $c_0^\star$-flux'' ---  in exactly the same 
manner  as in Regge pole models\cite{r4}, where it is called 
``the pomeron flux''.] $F_2^c(\beta,Q^2)$ is 
``the electromagnetic structure function of $c_0^\star$''
[the counterpart of $F_2(x_B,Q^2)$ of the proton] which
--- in analogy to proton (or any other hadron) ---
can be expressed as
\begin{equation}
\label{eee2}
\frac{F_2^c(\beta,Q^2)}{\beta}
= \sum_i e_i^2 [q_i^c(\beta,Q^2)+\bar q_i^c(\beta,Q^2)],
\end{equation}
where $q_i^c(\bar q_i^c)$ stands for the probability 
density for $\gamma^\star$ 
to interact with a quark (antiquark) of flavor $i$ and electric
charge $e_i$ which carries a fraction $\beta$ of the energy 
(or longitudinal momentum)
of $c_0^\star$. It is clear that 
Eq.(6) should be valid for all $x_P$-values in this kinematical
region, that is, both the right- and the left-hand-side
of Eq.(6) should be independent of the energy (momentum) carried
by the ``hadron'' $c_0^\star$.

Hence, to find out experimentally whether the second question can be 
answered in the affirmative, we only need to check whether the
data are in agreement with the assumption
that $F_2^c(\beta , Q^2)$ prescribed by Eqs.(5) and (6) exists.
For such a test,
we take the existing
data\cite{r3} and plot $\log [F_2^{D(3)}(\beta, Q^2;x_P)/\beta]$ 
against $\log\beta$ for different $x_P$-values.
We note, under the assumption 
that the factorization shown in Eq.(5)
is valid, the $\beta$-dependence for a given $Q^2$ in
such a plot should have exactly the same form as that in the  
corresponding
$\log [F_2^{c}(\beta, Q^2)/\beta]$ vs $\log \beta$ plot;
and that the latter is the analog of
$\log [F_2(x_B, Q^2)/x_B]$ vs $\log x_B$ plot for normal events. 
In Fig.2 we show the result of such
plots for three fixed $Q^2$-values (3.5, 20 and 65 GeV$^2$,
as representatives of three different ranges in $Q^2$). 
Our goal is to examine whether or
how the $\beta$-dependence of the function given in
Eq.(6) changes with $x_P$. In principle,
if there were enough data points, we should, and we could, do such
a plot for the data-sets associated with every $x_P$-value.
But, unfortunately there are not so much data at present.
What we can do, however, is to consider
the $\beta$-distributions in different $x_P$-bins, and to vary
the bin-size of $x_P$,
so that we can explicitly 
see whether/how the shapes of the $\beta$-distributions
change. The results are shown  
in Fig.2. The $\beta$-distribution in the first 
row, corresponds to the integrated value $\tilde{F}^D_2(\beta, Q^2)$
shown in the literature\cite{r3,r5}.
Those in the second and in the third row are obtained by considering
different bins and/or by
varying the sizes of the bins.
By joining the points associated with a given $x_P$-interval
in a plot for a given $Q^2$,
we obtain the $\beta$-distribution for a $c_0^\star$ carrying 
approximately the amount of energy $x_P P^0$, encountered 
by a photon of virtuality $Q^2$. Taken together with Eq.(6) we can 
then extract the distributions $q_i^c(\beta, Q^2)$ and
 $\bar{q}_i^c(\beta, Q^2)$  for this $Q^2$-value, provided
that $F_2^c(\beta, Q^2)/\beta$ is independent of $x_P$.
But, as we can see in Fig.2, the existing data\cite{r3,r5}
show that the $x_P$-dependence of this function is far from 
being negligible!
Note in particular 
that according to Eq.(\ref{eee1}), by choosing a suitable $f_P(x_P)$ 
we can shift the curves for different $x_P$-values in the vertical 
direction (in this log-log plot); but {\em we can never change 
the shapes of the $\beta$-distributions} which are different for
different $x_P$-values!  

In order to see, and to realize, the meaning of the $x_P$-dependence
of the distributions of the charged constituents of $c^\star_0$ 
expressed in terms of  $F_2^c(\beta, Q^2)/\beta$ 
in LRG events [see Eqs.(5) and (6)], 
let us, for a moment, consider
normal deep-inelastic scattering events in the 
$x_B$-region where quarks dominate ($x_B > 0.1$, say).
Here we can plot the data for 
$\log [F_2(x_B, Q^2)/x_B]$ as a function of $\log x_B$ obtained
at {\em different incident energies ($P^0$'s)} of the proton.
{\em Suppose} we see, that 
at a given $Q^2$, the data for $x_B$-distributions taken
at different values
of $P^0$ are very much different.
{\em Would} it still be possible to introduce $F_2(x_B,Q^2)$
as ``the electromagnetic  structure function'' of the proton,
from which we can extract the $x_B$-distribution of the quarks
$q_i(x_B,Q^2)$ at a given $Q^2$?

\subsection*{IV. Distributions of the  gluon-clusters}
After having seen that the existing data 
are not in agreement with the picture in which
the colorless gluon-clusters ($c_0^\star$'s) are 
hadron-like, we now come back 
to the first question in Section III, and try to find out whether it is
never-the-less possible and meaningful to talk about the 
$x_P$-distribution of $c_0^\star$. We shall see in this section,
the answer to this question is Yes! 
Furthermore, we shall also see,
in order to answer this question in the 
affirmative, we do not need the factorization mentioned 
in Eq.(5); and we do not need to know whether the gluon-clusters are
hadron-like. But, as we shall show later on, it is of considerable importance
to discuss the second question in understanding
the nature of the $c_0^\star$'s. 

In view of the fact that we do use the concept ``distributions
of gluons'' 
in deep-inelastic lepton-hadron scattering, although the gluons  
do not directly interact with the virtual photons,
we  shall try to introduce the notion ``distribution of 
gluon-clusters'' in a similar manner. 
In order to see what we should do for the introduction
of such distributions, let us recall the following:

For normal deep-inelastic $e^-p$ collision
events, the structure function $F_2(x_B, Q^2)$ can be expressed
in term of the distributions of partons, where the partons are 
not only quarks and antiquarks, but also gluons which 
can contribute to the structure function by quark-antiquark 
pair creation and annihilation. 
In fact, in order to satisfy energy-momentum-conservation
(in the electron-proton system),
the contribution of the gluons $x_gg(x_g,Q^2)$ has to be taken into account
in the energy-momentum sum rule
for all measured $Q^2$-values. Here, we denote by 
$g(x_g,Q^2)$ the probability density 
for the virtual photon $\gamma^\star$ (with virtuality $Q^2$) to meet a
gluon which carries the energy (momentum) fraction $x_g$ of the proton,
analogous to $q_i(x_B, Q^2)$ [or $\bar q_i(x_B, Q^2)$] which 
stands for the probability density for this $\gamma^\star$ 
to interact with a quark (or an antiquark) of flavor $i$ and electric 
charge $e_i$ which carries the energy (momentum) fraction $x_B$ of the 
proton. We note, while both $x_B$ and $x_g$ stand for energy 
(or longitudinal momentum) fractions carried by partons, 
the former can be, but the latter {\em cannot} be directly
measured. 

Having these, in particular the energy-momentum sum rule in mind,
we immediately see the following: In a given
kinematical region
in which the contributions of only
one category of partons (for example quarks for $x_B > 0.1$ or
gluons for $x_B < 10^{-2}$) dominate, the structure
function $F_2(x_B,Q^2)$ can approximately
be related to the 
distributions of that particular kind of partons in a
very simply manner. In fact,
the expressions below can be, and have been,
interpreted as the probability-densities for the virtual photon $\gamma^\star$
(with virtuality $Q^2$) to meet a quark or a gluon which carries the energy
(momentum) fraction $x_B$ or $x_g$ respectively. 

\begin{eqnarray}
\label{ee2}
{F_2(x_B,Q^2)\over x_B}\approx \sum_i e_i^2\, q_i(x_B,Q^2) &
\mbox{\hspace*{1cm}or\hspace*{1cm}} &
{F_2(x_B,Q^2)\over x_g}\approx g(x_g,Q^2)\mbox{\ .}
\end{eqnarray}
The relationship between $q_i(x_B,Q^2)$,
$g(x_g,Q^2)$ and
$F_2(x_B, Q^2)$ as they stand  in Eq.(\ref{ee2})
are general
and formal (this is the case especially for that between $g$ and
$F_2$) in the following sense: 
Both $q_i(x_B, Q^2)$ and $g(x_g,Q^2)$ contribute to the
energy-momentum sum rule and both of them are in accordance
with 
the assumption that partons 
of a given category
(quarks or gluons)
dominate a given kinematical region
(here $x_B>0.1$ and $x_B<10^{-2}$ respectively).
But, neither the dynamics which leads to the observed $Q^2$-dependence
nor the relationship between $x_g$ and $x_B$ are given. This means,
{\it without further theoretical inputs}, the simple expression for
$g(x_g, Q^2)$ as given by Eq.(7) is {\it practically useless}!

Having learned this, we now discuss what happens
if we assume, in diffractive lepton-nucleon scattering,
the colorless gluon-clusters ($c_0^\star$'s) dominate the 
small-$x_B$ region ($x_B< 10^{-2}$, say). In this simple picture, we
are assuming that the following is approximately true: 
The gluons in this region appear predominately in form
of gluon clusters. The interaction
between  the struck $c_0^\star$
and the rest of the proton
can be neglected during the 
$\gamma$-$c_0^\star$ collision such that
we can apply impuls-approximation to the $c_0^\star$'s in this
kinematical region.  
That is, here we can
introduce 
--- in the same manner as we do for
other partons
(see Eq.\ref{ee2}), a
probability density $D_S(x_P|\beta,Q^2)$ for $\gamma^\star$ in the 
diffractive scattering process to ``meet'' a $c_0^\star$ which carries 
the fraction $x_P$ of the proton's 
energy $P^0=(|\vec{P}|^2+M^2)^{1/2} \approx |\vec{P}|$
(where $\vec{P}$ is the momentum and $M$ is the mass of the proton).
In other words, 
in 
diffractive scattering events
for processes in the  kinematical region
$x_B < 10^{-2}$, we should have, instead of $g(x_g,Q^2)$, the following:
\begin{equation}
\label{ee3}
{F_2^{D(3)}(\beta,Q^2;x_P)\over x_P}\approx D_S(x_P|\beta,Q^2)\,.
\end{equation}
Here, $x_PP^0$ is the energy carried by $c_0^\star$, 
and $\beta$ indicates the corresponding fraction carried by
the struck charged constituent in $c_0^\star$.
In connection with the similarities and the differences between 
$q_i(x_B,Q^2)$, $g(x_B,Q^2)$ in (\ref{ee2}) and $D_S(x_P|\beta, Q^2)$
in (\ref{ee3}), it is
useful to note in particular the significant difference 
between $x_g$ and $x_P$, 
and thus that between 
the $x_g$-distribution $g(x_g,Q^2)$ of the gluons and the
$x_P$-distribution $D_S(x_P|\beta, Q^2)$ of the $c^\star_0$'s: Both $x_g$ and 
$x_P$ are energy (longitudinal momentum) fractions of charge-neutral 
objects, with which
$\gamma^\star$ {\it cannot} directly interact. But, in contrast to $x_g$,
$x_P$ {\it can be directly measured in experiments}, namely by 
making use of the kinematical relation
\begin{equation}
\label{ee4}
x_P\approx {Q^2+M_x^2\over Q^2+W^2},
\end{equation}
and
by measuring the quantities $Q^2$, $M_x^2$ and $W^2$
in every collision event. Here, $Q$, $M_x$
and $W$ stand respectively for the invariant momentum-transfer from
the incident electron, the invariant-mass of the final hadronic state
after the $\gamma^\star-c_0^\star$ collision, and the invariant mass of the
entire hadronic system in the collision between $\gamma^\star$ and the
proton. Note that $x_B\equiv\beta x_P$, hence $\beta$ is also
measurable. This means, in sharp contrast to $g(x_g,Q^2)$, {\it
experimental information} on $D_S(x_P|\beta, Q^2)$ 
in particular its $x_P$-dependence
can be obtained ---
{\it without further theoretical inputs}!
 
\subsection*{V. The first SOC-fingerprint: Spatial scaling}
We mentioned at the beginning of Section III, that in order
to find out whether the concept of SOC
indeed plays a role in diffractive DIS we need to check the
fingerprints of SOC shown in Section II, and that such tests 
can be made by examing the 
corresponding cluster-distributions obtained from experimental data.
We are now ready to do this, because we have
learned in Sections III and IV, that it is not only
meaningful but also possible to extract $x_P$-distributions from the 
measured diffractive structure functions,
although the gluon-clusters {\em cannot} be treated as hadrons.
In fact, as we can explicitly see
in Eqs.(8) and (9), in order to extract the $x_P$-dependence of
the gluon-clusters from the data, detailed knowledge about the intrinsic
structure of the clusters are not necessary.

Having these in mind, we now
consider $D_S$ as a function of $x_P$ for given values
of $\beta$ and $Q^2$,
and plot 
$F_2^{D(3)}(\beta,Q^2;x_P)/x_P$ against $x_P$
for different sets of $\beta$ and $Q^2$. The results of such  
log-log plots are shown in Fig. 3.  As we can see,
the data\cite{r3} suggest
that the probability-density for the virtual photon $\gamma^\star$  to 
meet a color-neutral and charged-neutral object $c_0^\star$ with energy
(longitudinal momentum) fraction $x_P$ has a power-law behavior in
$x_P$, and the exponent of this power-law
depends very little on $Q^2$ and $\beta$.  
This is to be compared with $D_S(S)$ in Eq.(~\ref{e1}), where $S$,
the dissipative energy (the size of the BTW-cluster)
corresponds to the energy of the
system $c_0^\star$. The latter is $x_PP^0$, where $P^0$ is the
total energy of the proton.

It means, the existing data\cite{r3} show that
$D_S(x_P | \beta, Q^2)$ exhibits the same kind of power-law behavior
as the size-distribution of BTW-clusters.
This result is 
in accordance with the expectation that
self-organized critical phenomena may exist
in the colorless systems of interacting soft gluons in
diffractive deep-inelastic electron-proton 
scattering processes.

We note, up to now, we have only argued (in Section I) that
such gluon-systems are
open, dynamical, complex systems
in which SOC may occur, and we have mentioned (in Section II) the 
ubiquitousness of SOC in Nature.
Having seen the first piece of experimental evidence
that one of the necessary conditions for the existence of
SOC is satisfied, let us now take a second look at the colorless 
gluon-systems from a theoretical point of view: Viewed from a 
``fast moving frame'' which can
for example be the electron-proton c.m.s. frame,  
such 
colorless systems
of interacting soft gluons are part of the proton
(although, as color-singlets, they can also be outside the confinement
region). Soft gluons can be
intermittently emitted or absorbed by gluons in such a system, 
as well as by gluons, 
quarks and antiquarks outside the system. 
The emission- and absorption-processes are due to local interactions
prescribed by the well-known QCD-Lagrangian (here ``the running
coupling constants'' are in general large,
because the distances between the interacting colored objects
cannot be considered as ``short''; remember that the 
spatial dimension of a $c_0^\star$ can be much
larger than that of a hadron!). 
In this connection, it is however very useful to keep in mind that,
due to the complexity of the system,
details about the local interactions may be relatively
unimportant, while
general and/or global features --- for example 
energy-flow between different parts (neighbors and neighbor's
neighbors $\ldots$) of the system --- 
may play an important role. 

How far can one go in neglecting dynamical details when one 
deals with such open 
complex systems? In order to see this, let us
recall how Bak and Sneppen\cite{r14}
succeeded in modelling
some of the essential aspects of 
The Evolution in Nature.
They consider the ``fitness'' of different ``species'', related to one
another through a ``food chain'', and assumed
that the species with the lowest fitness
is most likely to disappear or mutate at the next time-step
in their computer simulations. 
The crucial step in their simulations
that {\em drives} evolution is the adaption of the individual species to
its present {\em environment} (neighborhood) through mutation 
and selection of a 
fitter variant.
Other interacting species form part of the {\em environment}.
This means, the neighbors will be influenced by
every time-step.
The result these authors
obtained strongly suggests
that the process of evolution is
a self-organized critical phenomenon. One of the essential
simplifications they made in their evolution models\cite{r14,r15}
is the following: Instead of the explicit 
connection between
the fitness and the configuration of the
genetic codes,
they use random numbers for the fitness of the
species. 
Furthermore, as they have pointed out in their papers, they
 could in principle have chosen to model evolution on a less
coarse-grained scale by considering mutations at the individual
level rather than on the level of species, but that would make the 
computation prohibitively difficult.

Having these in mind, we are naturally led to the questions: 
Can we consider the creation and
annihilation processes of colorless
systems of interacting soft gluons associated
with a proton as ``evolution'' in a microscopic world?
Before we try to build models for a quantitative description
of the data, can we simply apply the existing evolution
models\cite{r14,r15} to such open, dynamical, complex 
systems of interacting soft-gluons,
and check whether some of the essential features 
of such systems
can be
reproduced?

To answer these questions, we now report on the result of our 
first trial in this direction: 
Based on the fact that we know {\em very little} about
the detailed reaction mechanisms in such gluon-systems and 
{\em practically}
{\em nothing} about their structures, we simply {\em ignor} them, 
and assume that they are self-similar in space
(this means, colorless gluon-clusters can be considered as clusters of
colorless gluon-clusters and so on). Next,
we divide them in an arbitrary given number of subsystems $s_i$
(which may or may not have the same size). Such a system is open,
in the sense that neither its energy $\varepsilon_i$, nor
its gluon-number $n_i$ has a fixed value. Since we do not
know, in particular, how large the $\varepsilon_i$'s are, we  
use random numbers. As far the $n_i$'s are concerned, since
we do not know how these numbers are associated with the energies
in the subsystems $s_i$, except that they are not conserved
quantities,
we just ignor them, and consider only the $\varepsilon_i$'s.
As in Ref.[\ref{r14}] or in Ref.[\ref{r15}], the random number of this
subsystem as well as those of the fixed\cite{r14} or random (see the
first paper of Ref.[\ref{r15}]) neighbors will be changed at every time-step.
Note, this is how we simulate the processes of energy flow due to 
exchange of gluons between the subsystems, as well as those with
gluons/quarks/antiquarks outside the system. In other words, in the
spirit of Bak and Sneppen\cite{r14} we neglecting the dynamical
details {\it totally}.
Having in mind that, 
in such systems,
the gluons as well as the
subsystems ($s_i$'s) of gluons  are {\it virtual}
(space-like), we can ask: 
``How long can such a colorless subsystem
$s_i$ of interacting soft gluons exist,
which carries energy $\varepsilon_i$?''
According to the uncertainty principle, 
the answer should be:
``The time interval
in which the subsystem $s_i$ can exist
is proportional to $1/\varepsilon_i$,
and this quantity can be considered as the lifetime $\tau_i$ of
$s_i$.'' In this sense, the subsystems of colorless gluons are
expected to have larger probabilities to mutate because they are
associated with higher energies, and thus shorter ``lifetimes''.
Note that the basic local interaction
in this self-organized evolution
process is the emission (or absorption) of gluons by gluons prescribed
by the QCD-Lagrangian --- although the detailed mechanisms 
(which can in principle be explicitly written down by
using the QCD-Lagrangian)
do not play a
significant role. 

In terms of the evolution model\cite{r14,r15}
we now call $s_i$ the ``species'' and identify 
the corresponding
lifetime $\tau_i$ as the ``fitness of $s_i$''.
Because of the one-to-one correspondence between $\tau_i$ and 
$\varepsilon_i$, where the latter is a random number,
we can also directly assign random numbers to the $\tau_i$'s
instead. From now we can adopt the evolution model\cite{r14,r15}
and note that,
at the start of such a process (a simulation), the fitness on average
grow, because the least fit are always eliminated. Eventually the
fitness do not grow any further on average. All gluons have a fitness
above some threshold. At the next step, the least fit species (i.e. the
most energetic subsystem $s_i$ of interacting soft gluons), 
which would be right at the threshold, 
will be ``replaced''
and starts an
avalanche (or punctuation of mutation events), which is causally
connected with this triggering ``replacement''. 
After a while, the avalanche will
stop, when all the fitnesses again will be over that threshold. 
In this sense, the evolution goes on, and on, and on.
As in Refs.[\ref{r14}] and [\ref{r15}], we can monitor the duration of
every avalanche, that is the total number of mutation events in
everyone of them, and count how many avalanches of each size are observed.
The
avalanches mentioned here are 
special cases of those discussed in Section II.
Their size- and lifetime-distributions are
given by Eq.(1) and Eq.(2) respectively. Note in particular that the
avalanches in the Bak-Sneppen model correspond to sets of subsystems
$s_i$, the energies ($\epsilon_i$) of which are too high ``to be fit
for the colorless systems of low-energy gluons''. It means, in the
proposed picture, what the virtual photon in deep-inelastic
electron-proton scattering ``meet'' are those ``less fit'' one ---
those who carry ``too much'' energy. 
In a geometrical picture this means, it is 
more probable for such ``relatively energetic'' colorless
gluons-clusters to be spatially
further away from the (confinement region of)
the proton.

There exists, in the mean time, already several versions of evolution 
models\cite{r10,r15} based
on the original idea of Bak and Sneppen\cite{r14}
Although SOC phenomena have been observed in all these cases\cite{r10,r14,r15},
the slopes of the power-law distributions for the avalanches are different
in different models --- depending on the rules applied to the mutations. 
The values range from  approximately $-1$ to approximately $-2$.
Furthermore, these models\cite{r10,r14,r15} seem to show that neither the
size nor the dimension of the system used for the computer simulation
plays a significant role.

Hence, if we identify 
the colorless charge-neutral object $c_0^\star$ encountered by the
virtual photon $\gamma^\star$ with 
such an avalanche,
we are identifying the
lifetime of $c_0^\star$ with $T$, and the ``size''
(that is the total amount of dissipative energy in this
``avalanche'') with the total amount of energy of $c_0^\star$.
Note that the latter is nothing else but $x_PP^0$, where $P^0$
is the total energy of the proton. This is how and why the
$S$-distribution in Eq. (\ref{e1}) and the $x_P$-distribution of
$D_S(x_P|\beta,Q^2)$  in Eq.(\ref{ee3}) are related to each other.

\subsection*{VI. The second fingerprint: Temporal scaling}
In this section we discuss in more detail the effects associated with 
the time-degree-of-freedom. In connection with the two questions
raised in Section III,
one may
wish to know
{\em why} the parton-picture 
does  not {\em always} work when we apply it in a straightforward 
manner --- not only to hadrons but also to gluon-clusters.
The answer is very simple: 
The time-degree
of freedom cannot be ignored when we wish to find out whether
impulse-approximation
is applicable, and the applicability of the latter
is the basis of the parton-model.
We recall that,
when we apply this model to stable hadrons,
the quarks, antiquarks and
gluons are considered as free and stable objects, 
while the virtual photon $\gamma^\star$ is associated
with a given interaction-time $\tau_{\rm int}(Q^2,x_B)$ characterized
by the values $Q^2$ and $x_B$ of such scattering processes. 
We note however that,  this is possible only when the interaction-time
$\tau_{\rm int}$ is much shorter than the 
corresponding 
time-scales (in particular the average
propagation-time of color-interactions in hadron).
Having these in mind, we see that, we are confronted with
the following questions when we  deal
with gluon-clusters associated with finite lifetimes:
Can we consider the $c_0^\star$'s as ``{\it free}'' and 
``{\it stable}'' particles when
their lifetimes are {\it shorter} than the interaction-time $\tau_{\rm
int}(Q^2,x_B)$? Can we say that a $\gamma^\star-c_0^\star$
collision process takes place,
in which the incident $\gamma^\star$ is
absorbed by one a or a system of the charged constituents of $c_0^\star$, when
the lifetime $T$ of $c_0^\star$ is {\it shorter} than 
$\tau_{\rm int}(Q^2,x_B)$?

Since the notion
``stable objects'' or ``unstable objects'' depends on the 
scale which is used in
the measurement, the question whether a $c_0^\star$ can 
be considered as a parton
(in the sense that it can be considered as a free 
``stable object'' during the $\gamma^\star$-$c_0^\star$ interaction)
depends very much on 
on the interaction-time
$\tau_{int}(Q^2, x_B)$. 
Here, for
given values of $Q^2$, $x_B$, and thus $\tau_{int}(Q^2, x_B)$,
only those $c^\star_0$'s whose lifetime ($T$'s) are greater
than $\tau_{int}(Q^2, x_B)$ can absorb the corresponding
$\gamma^\star$.
That
is to say, when we consider diffractive electron-proton scattering in
kinematical regions in which $c_0^\star$'s dominate, 
we must keep in mind that the measured cross-sections (and thus the 
diffractive structure function $F_2^{D(3)}$)
only include contributions from collision-events in which the
condition $T>\tau_{\rm int}(Q^2,x_B)$ 
is satisfied\,! 

We note that $\tau_{\rm int}$ can be estimated by making use of the 
uncertainty principle. In fact, by  calculating $1/q^0$ in the 
above-mentioned 
reference frame, 
we obtain
 \begin{equation}
\label{e4}
\tau_{\rm int}={4|\vec P|\over Q^2}{x_B\over 1-x_B},
\end{equation}
which implies that, for given $|\vec P|$ and $Q^2$ values, 
\begin{equation}
\label{eee3}
\tau_{\rm int}\propto x_B,\hskip 1cm \mbox{\rm for } x_B\ll 1.
\end{equation} 
This means, for diffractive $e^-p$ scattering events in the
small-$x_B$ region at given $|\vec
P|$ and $Q^2$ values, $x_B$ is directly proportional to the interaction
time $\tau_{\rm int}$. Taken together with the relationship between
$\tau_{\rm int}$ and the minimum lifetime $T({\rm mim})$ of the
$c_0^\star$'s mentioned above, we reach the following conclusion: The
distribution of this minimum value,
$T({\rm min})$ of the $c_0^\star$'s which dominate the
small-$x_B$ ($x_B<10^{-2}$, say) region can be obtained by examining
the $x_B$-dependence of 
$F_2^{D(3)}(\beta,Q^2;x_P)/\beta$ discussed in
Eqs. (5), (6) and in Fig. 2. This is because, due to the fact that 
this function is proportional to
the quark (antiquark) distributions $q^c_i(\bar{q_i}^c)$ which can be 
directly probed by the incident virtual photon
$\gamma^\star$, by measuring $F_2^{D(3)}(\beta,Q^2,x_P)/\beta$
as a function of $x_B\equiv \beta x_P$, we are in fact asking
the following questions:
Do the distributions of the charged constituents of
$c_0^\star$ depend on the interaction time $\tau_{\rm int}$,
and thus on the minimum lifetime  $T({\rm min})$ of the 
to be detected gluon-clusters\,?
We use the identity $x_B\equiv\beta x_P$ and plot the quantity
$F_2^{D(3)}(\beta,Q^2;x_P)/\beta$ against the variable $x_B$
for fixed values of $\beta$ and $Q^2$.
The result of such a log-log plot is given in Fig.4. It shows
not only  how the dependence on the
time-degree-of-freedom can be extracted from the existing
data\cite{r3}, but also that, for all the measured  
values of $\beta$ and $Q^2$, the quantity
\begin{equation}
\label{e5}
 p(x_B|\beta, Q^2) \equiv 
{F_2^{D(3)}(\beta, Q^2; x_B/\beta)
\over \beta }
\end{equation}
is approximately
independent of $\beta$, and independent of $Q^2$.
This strongly suggests that the quantity given in Eq.(\ref{e5})
is associated with some {\em global} features of $c_0^\star$ ---
consistent with the observation made in Section III which shows
that it cannot be used to describe the structure of $c_0^\star$.
This piece of empirical fact can be expressed by setting
$p(x_B|\beta, Q^2)\approx p(x_B)$. 
By taking a closer look at this $\log$-$\log$ plot, as well
as the corresponding plots for different sets of
fixed $\beta$- and $Q^2$-values (such plots are not
shown here, they are similar to those in Fig.3),
we see that they are straight lines indicating that 
$p(x_B)$ obeys a power-law. What does this piece of 
experimental fact tell us? What can we learn from
the distribution of the lower limit of the lifetimes (of the
gluon-systems $c_0^\star$'s)? 

In order to answer these questions, let us,
for a moment, assume that we know the lifetime-distribution $D_T(T)$
of the $c_0^\star$'s. In such a case,
 we can readily evaluate the  integral
\begin{equation}
\label{e6}
I[\tau_{\rm int}(x_B)]\equiv\int^\infty_{\tau_{\rm int}(x_B)}D_T(T)dT,
\end{equation}
and thus obtain the number density of all those clusters which live 
longer than the interaction time $\tau_{\rm int}(x_B)$.
Hence, under the statistical assumption 
that the chance for a $\gamma^\star$
to be absorbed by one of those
$c_0^\star$'s of lifetime $T$
is proportional to $D_T(T)$ (provided that
$\tau_{\rm int}(Q^2,x_B)\le T$, otherwise this chance is 
zero), we 
can then interpret the integral in Eq.(13) as follows:
$I[\tau_{\rm int}(Q^2,x_B)]\propto p(x_B)$ is
the probability density for $\gamma^\star$ [associated with the
interaction-time $\tau_{\rm int}(x_B)$] 
to be absorbed by $c_0^\star$'s.
Hence,
\begin{equation}
\label{e7}
D_T(x_B)\propto {d\over dx_B}p(x_B).
\end{equation}
This means in particular,  
 the fact that $p(x_B)$ obeys a power-law in $x_B$ implies that
$D_T(T)$ obeys a  power-law in $T$.
Such a {\em behavior is similar} to 
that shown in Eq.(~\ref{e2}).
In order to see the {\em quality} of this power-law behavior of $D_T$, and
the {\em quality} of its independence of $Q^2$ and $\beta$, we compare the
above-mentioned behavior with the existing data\cite{r3}. In Fig.5,
we show the log-log plots of
$d/dx_B[p(x_B)]$ against $x_B$. We note that $d/dx_B[p(x_B)]$ is
approximately $F_2^{D(3)}(\beta, Q^2; x_B/\beta)/(\beta x_B)$.
The quality of the power-law 
behavior of $D_T$ is explicitly shown in Fig.5.

\subsection*{VII. $Q^2$-dependent exponents in the power-laws?}
We have seen, in Sections V and VI, that in 
diffractive deep-inelastic
electron-proton scattering, the size- and the 
lifetime-distributions of the gluon-clusters obey power-laws,
and that the exponents depend very little
on the variables $\beta$ and $Q^2$. We interpreted
the power-law behaviors as the fingerprints of SOC in the formation
processes of such clusters. Can such approximately
independence (or weak
dependence) of the exponents on $Q^2$ and $\beta$ 
be understood in a physical picture based on
SOC? In
particular, what do we expect to see in photoproduction 
processes
where the
associated value for $Q^2$ is zero?

In order to answer these questions, let us 
recall the space-time aspects of the
collision processes which are closely related
to the above-mentioned
power-law behaviors.
Viewed in a fast moving frame (e.g. the c.m.s. of the colliding
electron and proton), the states of the interacting soft gluons 
originating from  the
proton are self-organized.
The colorless gluon-clusters caused by local perturbations
and developed through ``domino effects'' are BTW-clusters.
That is, they are avalanches (see Sections I and V), the 
size-distribution of which [see Eqs.(8) and (1)] are given by
Fig.3. This explicitly shows that
there are gluon-clusters of all sizes,
because a power-law
size-distribution implies that there is no scale in size.
Recall that, since such clusters are color-singlets, their 
spatial extensions can be much larger than that of the proton,
and thus they can  be ``seen'' also {\em outside} the proton 
by a virtual
photon originating from the electron. 
In other words, what the virtual photon encounters is a cloud of 
colorless gluon-clusters
spatially extended in- and outside the proton.

The virtual photon, when it encounters a colorless
gluon-cluster, will be absorbed 
by the charged constituents
(quarks and antiquarks due to fluctuation of the gluons)
of the gluon-system. Here it is useful to recall that in such a space-time
picture, $Q^2$ is inversely proportional to the transverse size,
and $x_B$ is a measure of the interaction time [See Eqs. (10) and (11)
in Section VI] of the virtual photon.
It is conceivable, that the values for the cross-sections for virtual
photons (associated with a given $Q^2$ and a given $x_B$) to 
collide with gluon-clusters (of a given size and a given 
lifetime) may depend on these variables. But, since the
processes of self-organization (which produce such gluon-clusters) 
take
place independent of the virtual photon (which originates from the 
incident electron and enters ``the cloud'' to
look for suitable partners), the power-law behaviors of the size-
and lifetime-distributions of the gluon-clusters are expected to be
independent of the properties associated with the virtual photon.
This means, by using
$\gamma^\star$'s associated with different values 
of $Q^2$ to detect clusters of various sizes, 
we are moving up or down on the straight lines in the 
log-log plots for the size- and lifetime distributions, 
the slopes of which do not change.
In other words,
the approximative $Q^2$-independence of the slope is 
a natural consequence of the SOC picture.

As far as the $\beta$-dependence is concerned, we recall the
results obtained  in Sections III and IV,
which explicitly show the following:
The gluon-clusters ($c_0^\star$'s)
can {\em not} be considered as hadrons. In particular, it is
neither possible
nor meaningful
to talk about ``the electromagnetic structure of the gluon-cluster''. 
This suggests, by studying the $\beta$-dependence of
the ``diffractive structure functions'' we cannot expect to gain further
information about the structure
of the gluon-clusters or further insight about the reaction mechanisms. 

Having seen these, we try to look for
measurable quantities in which the integrations over $\beta$ 
have already been
carried out.
A suitable candidate for this purpose is the differential cross-section
\begin{eqnarray}
\frac{1}{x_P}\,\frac{d^2\sigma^D}{dQ^2 dx_P} 
& = &
\int d\beta \,\frac{4\pi\alpha^2}{\beta Q^4}\,
            \left( 1-y+\frac{y^2}{2}\right)\,
            \frac{F_2^{D(3)}(\beta, Q^2; x_P)}{x_P} \nonumber\\
& \approx &
\int d\beta \,\frac{4\pi\alpha^2}{\beta Q^4}\,
            \left( 1-y+\frac{y^2}{2}\right)\,
             D_S(x_P| \beta, Q^2)
\end{eqnarray}
Together with Eqs.(3) and (8), we see that this cross-section is 
nothing else but the effective $\beta$-weighted
$x_P$-distribution $D_S(x_P|Q^2,\beta)$ of the 
gluon-clusters. Note that the weighting factors shown on the 
right-hand-side of Eq.(15) are simply results of QED! 
Next, we use the data\cite{r3} for 
$F_2^{D(3)}$ which are available at present,
to do a log-log plot for the integrand of the expression
in Eq.(15) as a function of $x_P$
for different values of $\beta$ and $Q^2$.
This is shown in 
in Fig.6a. Since the absolute values of this quantity depend
very much, but the slope of the curves very little on $\beta$,
we carry out the integration as follows:
We first fit every set of the data separately.
Having obtained the slopes and the intersection points,
we use the obtained fits to perform the integration over $\beta$.
The results are shown in the
\begin{eqnarray*}
\log{\left(\frac{1}{x_P}\,\frac{d^2\sigma^D}{dQ^2\,dx_P}\right)}
& \mbox{\ \ versus\ \ \ } &
\log{(x_P)}
\end{eqnarray*}
plots of Fig.6b.
These results show the $Q^2$-dependence of the slopes
is practically negligible, and that the slope
is approximately $-1.95$ for all values of $Q^2$.

Furthermore, in order to see whether the quantity introduced in
Eq.(15) is indeed
useful, and in order to perform a decisive test of the 
$Q^2$-independence of the slope in the power-law behavior
of the above-mentioned size-distributions,
we now
compare the results in deep-inelastic
scattering\cite{r3} with those obtained in photoproduction\cite{r16},
where LRG events have
also be observed. This means, as in
diffractive deep-inelastic scattering, we again associate the
observed effects with colorless objects which are interpreted as
system of interacting soft gluons originating from the proton.
In order to find out whether it is the same kind of 
gluon-clusters as in deep-inelastic scattering, and whether
they ``look'' very much different when we probe them with
real ($Q^2=0$) photons, we replot the existing
$d\sigma/dM_X^2$ data\cite{r16} for photoproduction 
experiments performed at different total energies,
and note
the kinematical relationship between $M_X^2$, $W^2$ and $x_P$
for $Q^2\ll M^2$ and $|t|\ll M_X^2$:
\begin{eqnarray}
x_P \approx \frac{M_X^2 + t}{W^2 - M^2}\approx \frac{M_X^2}{W^2} & &
\end{eqnarray}
The result of the corresponding
\begin{eqnarray*}
\log{\left(\frac{1}{x_P}\,\frac{d\sigma}{dx_P}\right)}
& \mbox{\ \ versus\ \ \ } &
\log{(x_P)}
\end{eqnarray*}
plot is shown in Fig.7. The slope obtained from  a least-square
fit to the existing data\cite{r16} is $-1.98\pm 0.07$.

The results obtained in diffractive
deep-inelastic electron-proton scattering
and that for diffractive photoproduction strongly suggest
the following: The formation processes of gluon-clusters
in the proton is due to self-organized criticality, and thus
the spatial distributions of such clusters
--- represented by the $x_P$-distribution ---
obey power-laws.
The exponents of 
such power-laws are
independent of
$Q^2$. Since $1/Q^2$ can be interpreted,
in a geometrical picture, as a measure for the transverse 
size  of the incident virtual photon, the observed 
$Q^2$-independence of the exponents can be
considered as further evidence for SOC ---
in the sense that the self-organized gluon-cluster formation
processes take place independent of the virtual
photon (which is ``sent in'' to detect the clusters).

\subsection*{VIII Concluding remarks}
The existence of large rapidity gap (LRG) events\cite{r2,r3} in deep-inelastic
electron-proton scattering is one of the most striking
features, if not {\em the} most striking feature of
the experimental data obtained in the small-$x_B$ ($x_B<10^{-2}$, say)
region. Taken together with the empirical facts\cite{r1} 
that gluons dominate
in this kinematical region and that their interactions are not
negligible, it seems quite natural to think, that such events are due to
collisions between the virtual photons originated from
the lepton and colorless gluon-systems originating from the proton. 

What we propose in the present paper is {\it a statistical
approach} to study such colorless gluon-systems.
The reasons, why we think such an approach is useful, can be 
summarized as follows:

First, a number of theoretical arguments and experimental 
indications suggest that such
 a system of interacting soft-gluons
is a system with
the following properties:
(a) It is a complex system with many degrees of freedom,
    because in general it has a large --- unknown --- number of
    gluons.
(b) It is an open system. This is because the members of a  colorless 
    gluon-system may interact (through emission and/or absorption
    of soft gluons) not only with one another, but also with
    gluons and/or quarks and antiquarks outside the system. Thus,  
    due to such interactions, neither the gluon-number
    nor the energy of this system can remain constant.
(c) It is neither in chemical nor in thermal equilibrium.
    This is because, as we can for example see in the analysis shown
    in Section III, it is not possible to
    consider the colorless gluon-cluster $c_0^\star$
    as a hadron-like object which has a given structure. In this sense, 
    we are forced to consider it as
    a dynamical system --- probably very far from 
    thermal and chemical equilibria.    
(d) The basic interactions between the members of the system, as well as 
    those between a member and quarks or gluons outside the system, are local.
    In fact, they are explicitly given by the well-known QCD-Lagrangian.
    But, as it is often the case in complex systems, whether
    the local dynamical details or the general global features of
    the system plays a more significant role is a different
    question.

Second, it has been proposed by Bak, Tang and Wiesenfeld\cite{r9,r10} 
some time ago, 
that a wide class of open dynamical complex systems far from equilibrium 
may evolve in a self-organized manner to critical 
states, which give rise to
spatial and temporal power-law scaling behaviors.
Such scaling behaviors are universal and robust,
in fact they can be considered as the ``fingerprints'' of self-organized
criticality (SOC).
In the macroscopic world, there are many open dynamical
complex systems which show this kind of scaling 
behaviors\cite{r9,r10}. 
Under the condition
(see above) that the colorless system of interacting gluons 
can 
indeed
be considered as an open, dynamical, complex system, it would 
be of considerable interest to see whether there can be 
self-organized criticality 
also in the microscopic world --- at the level
of gluons and quarks.

Third, by using the existing data for
deep inelastic electron-proton scattering\cite{r3}
 and those for photoproduction\cite{r16},
where colorless systems of interacting soft-gluons are expected to play a 
dominating role, we checked the above-mentioned fingerprints.
The obtained results show that {\em the above-mentioned characteristic 
features for SOC indeed exist}. 
Furthermore, it is seen that the relevant exponents 
in such power-laws
are {\em the same
for different reactions}.
The existence of SOC in systems of interacting soft gluons in such
reactions has a number of  consequences. It seems worthwhile
to study them in more detail. In particular, it would be very helpful to
build realistic models and/or cellular automata to 
do quantitative calculations.

Fourth, based on the obtained results in
particular the validity of the power-law behaviors,
the physical picture for a colorless gluon-cluster 
should be
as follows:
It is {\em not a hadron} with a given
structure. It has {\em neither a typical size, nor a typical
lifetime}, and its structure is changing all the time.
In fact, it  has much in common with an earthquake or an avalanche
(mentioned in more detail in Sections II, IV and V).
Can we learn more about these objects by studying other
reactions\,? Can we use the same concepts and methods to treat
hadron-hadron and hadron-nucleus collision processes\,?
It is known that ``the exchange of colorless objects'' plays
an important role also in diffractive hadron-hadron collisions.
Shall we
see this kind of power-law behaviors also in 
diffractive inelastic hadron-hadron scattering processes?
Studies along this line are in progress. 
The results will be published elsewhere, when they are ready.

\subsection*{Acknowledgments}
We  thank  P. Bak, X. Cai, D. H. E. Gross, C. S. Lam, Z. Liang,
K. D. Schotte, K. Tabelow and E. Yen
for helpful discussions, R. C. Hwa, C. S. Lam and J. Pan for
correspondence, and FNK der FU-Berlin
for financial support (FPS ``cluster'' der FU Berlin).
Y. Zhang also thanks Alexander 
von Humboldt Stiftung for the fellowship granted to him.

\newpage

\begin{figure}\label{figure1}
\caption{
The well-known Feynman diagrams (a) for diffractive and (b) for normal
deep-inelastic 
electron-proton scattering are shown together with 
the relevant kinematical variables which describe
such processes.}
\end{figure}

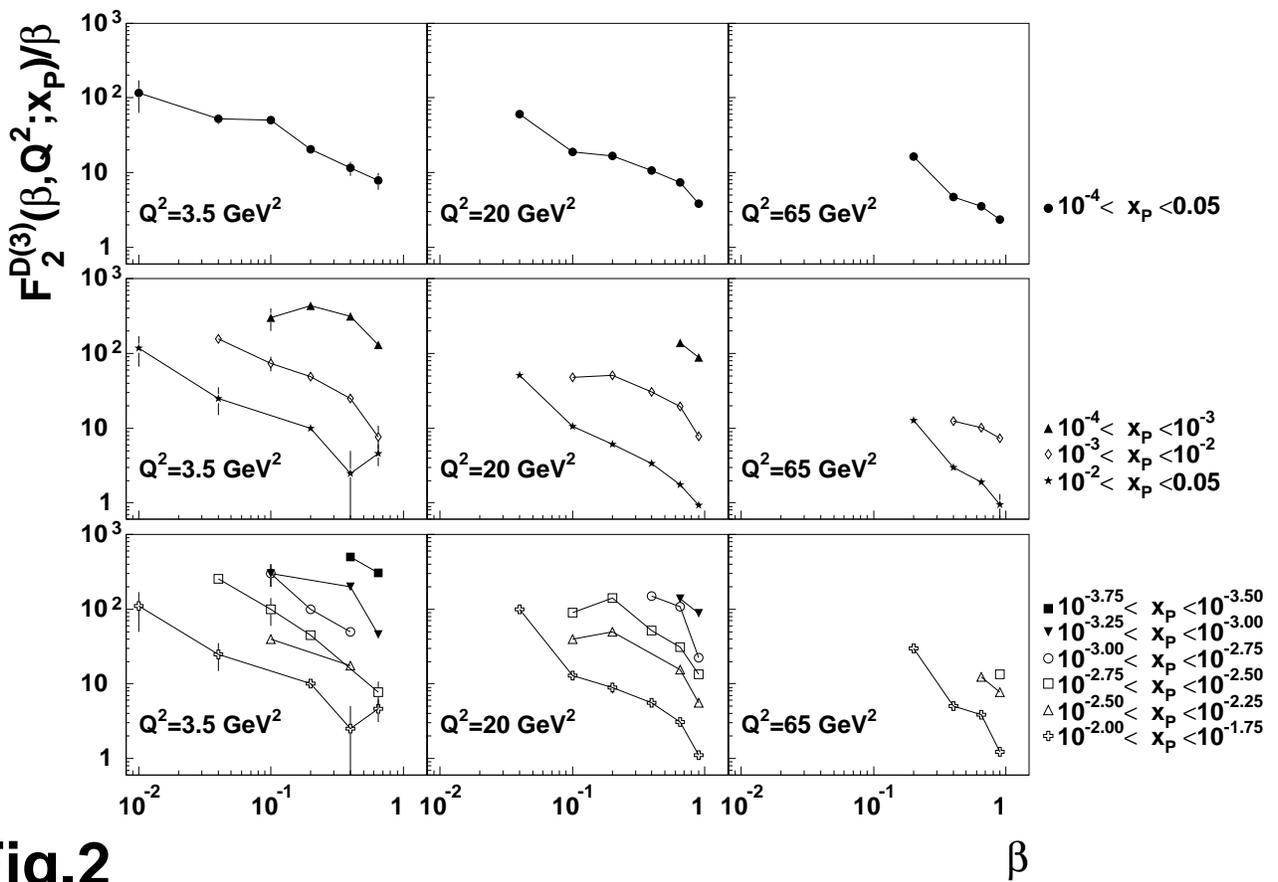
\begin{figure}\label{figure2}
\caption{$F_2^{D(3)}(\beta,Q^2;x_P)/\beta$ is plotted as a function of 
$\beta$ for given $x_P$-intervals and for fixed $Q^2$-values. 
The data are taken from Ref.[\ref{r3}].
The  lines  are only to guide the eye. }
\end{figure}

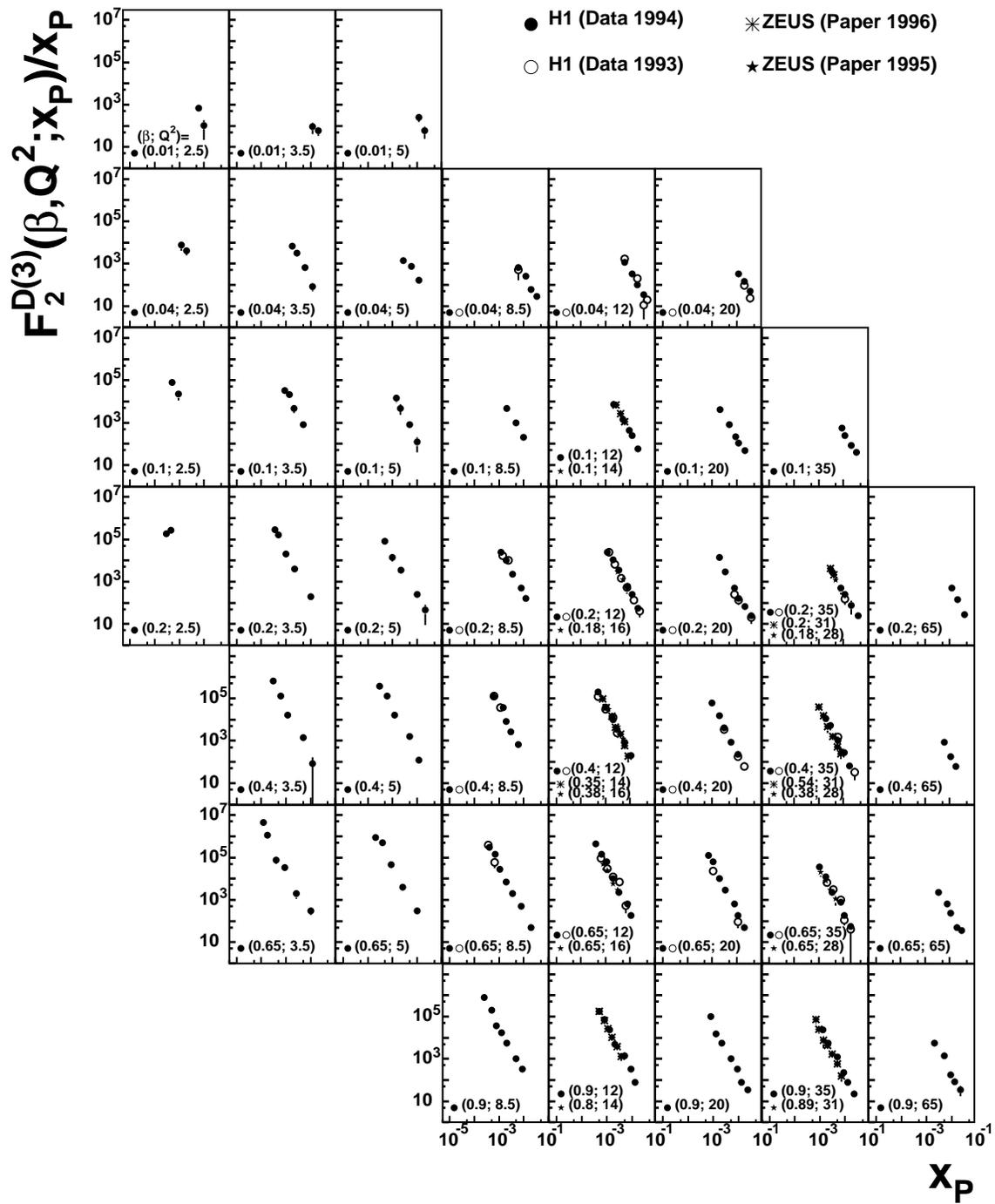
\begin{figure}\label{figure3}
\caption{$F_2^{D(3)}(\beta,Q^2;x_P)/x_P$ is plotted as a function of
$x_P$  for different values of $\beta$ and $Q^2$.  
The data are taken from 
Ref.[\ref{r3}].}
\end{figure}

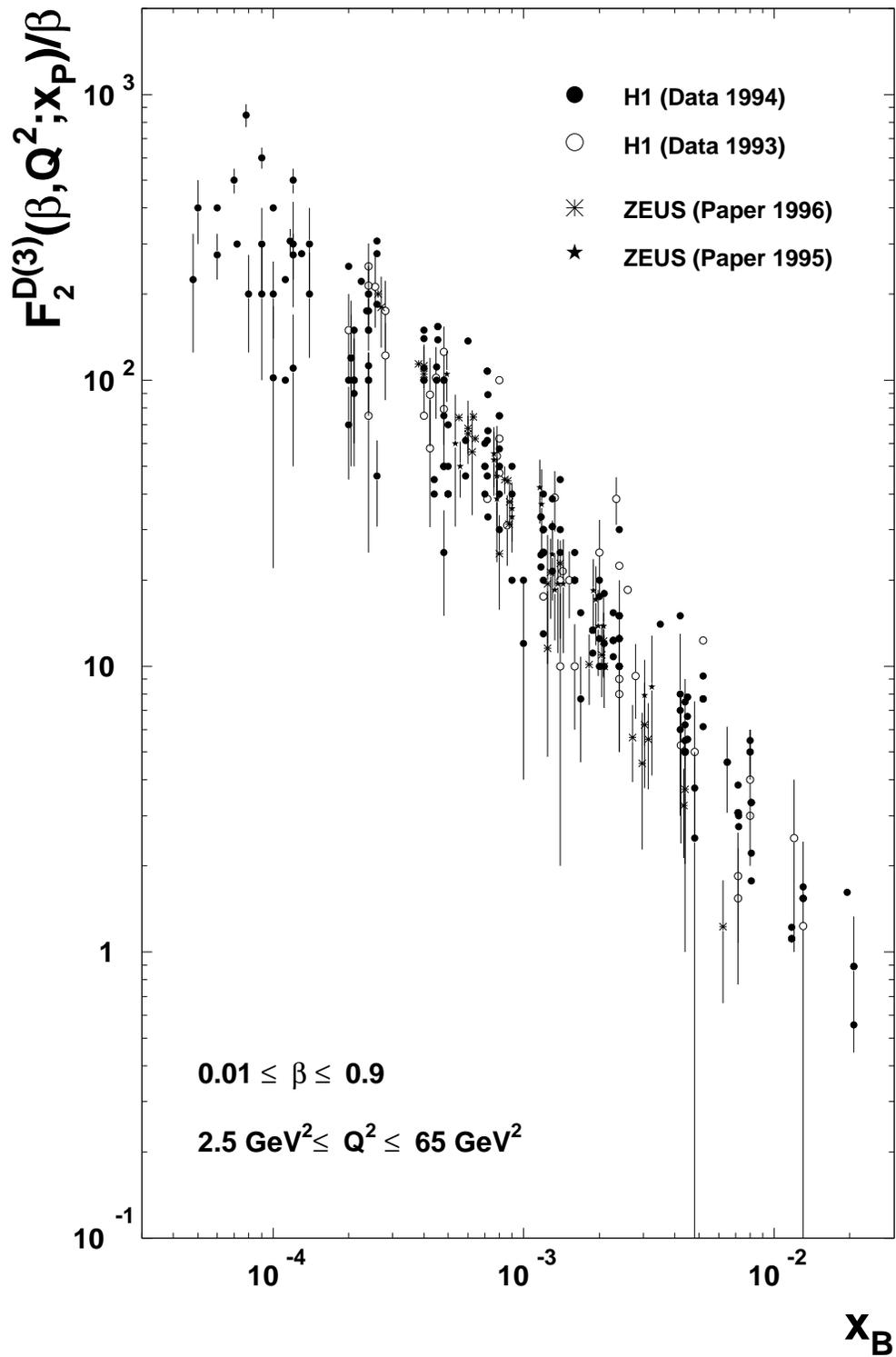
\begin{figure}\label{figure4}
\caption{$F_2^{D(3)}(\beta,Q^2;x_P)/\beta$ is plotted as a function of 
$x_B$ in the indicated $\beta$- and $Q^2$-ranges. 
The data are taken from Ref.[\ref{r3}].}
\end{figure}

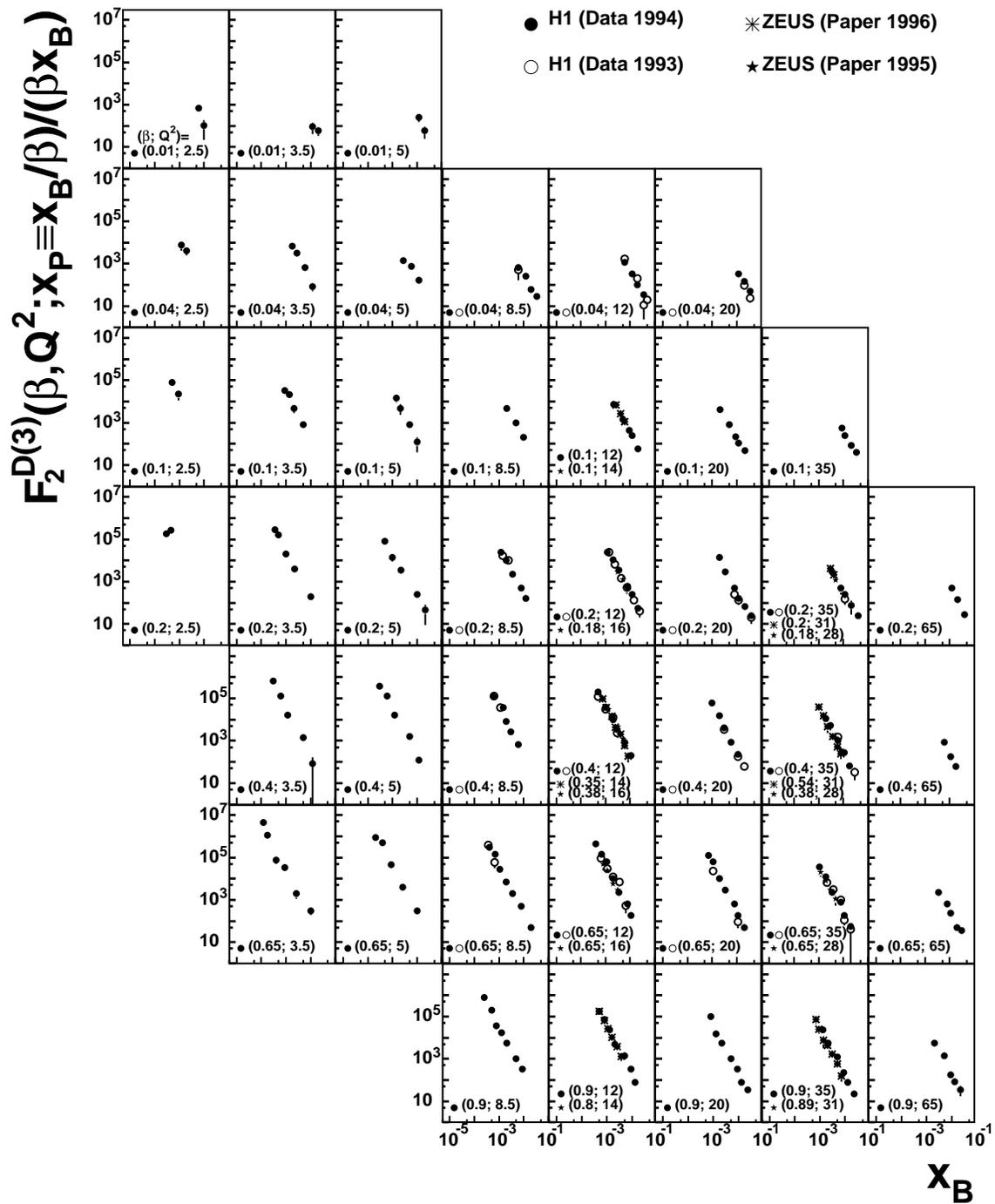
\begin{figure}\label{figure5}
\caption{$F_2^{D(3)}(\beta,Q^2;x_B/\beta)
/(\beta x_B)$ is plotted as a function of 
$x_B$ for fixed  $\beta$- and $Q^2$-values. 
The data are taken from Ref.[\ref{r3}].}
\end{figure}

\begin{figure}\label{figure6}
\caption{Figure 6: (a) $(1/x_P)d^3\sigma^D/d\beta dQ^2 dx_P$
            is plotted as a function of $x_P$ in different
            bins of $\beta$ and $Q^2$. The data are taken from
            Ref.[\ref{r3}].
         (b) $(1/x_P)d^2\sigma^D/ dQ^2 dx_P$
            is plotted as a function of $x_P$ in different
            bins of $Q^2$. The data are taken from
            Ref.[\ref{r3}].}
\end{figure}

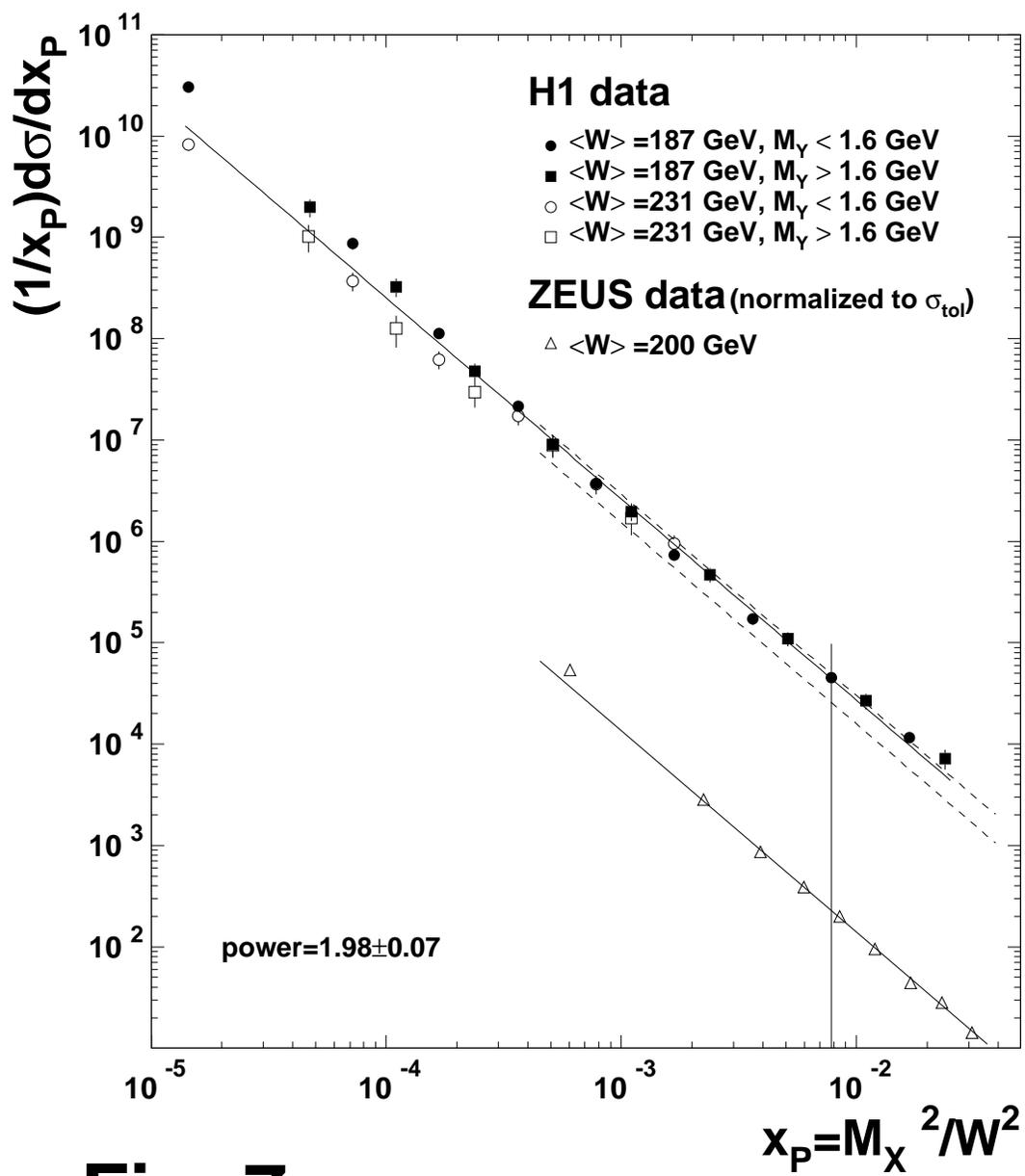
\begin{figure}\label{figure7}
\caption{ Figure 7: $(1/x_P)d\sigma/dx_P$ for photoproduction
          $\gamma + p \rightarrow X + p$ 
            is plotted as a function of $x_P$.
            The data are taken from Ref.[\ref{r16}].
           Note that the data in the second paper are given in terms of 
           relative cross sections. Note also that the slopes of the 
           straight-lines are the same. The two dashed lines indicate
           the lower and the upper limits of
           the results obtained by multiplying the lower solid line
           by $\sigma_{\mbox{tot}}=154\pm 16 \mbox{(stat.)} \pm
            32 \mbox{(syst.)} \mu b$. This value is taken from the third
           paper in Ref.[\ref{r16}].}
\end{figure}

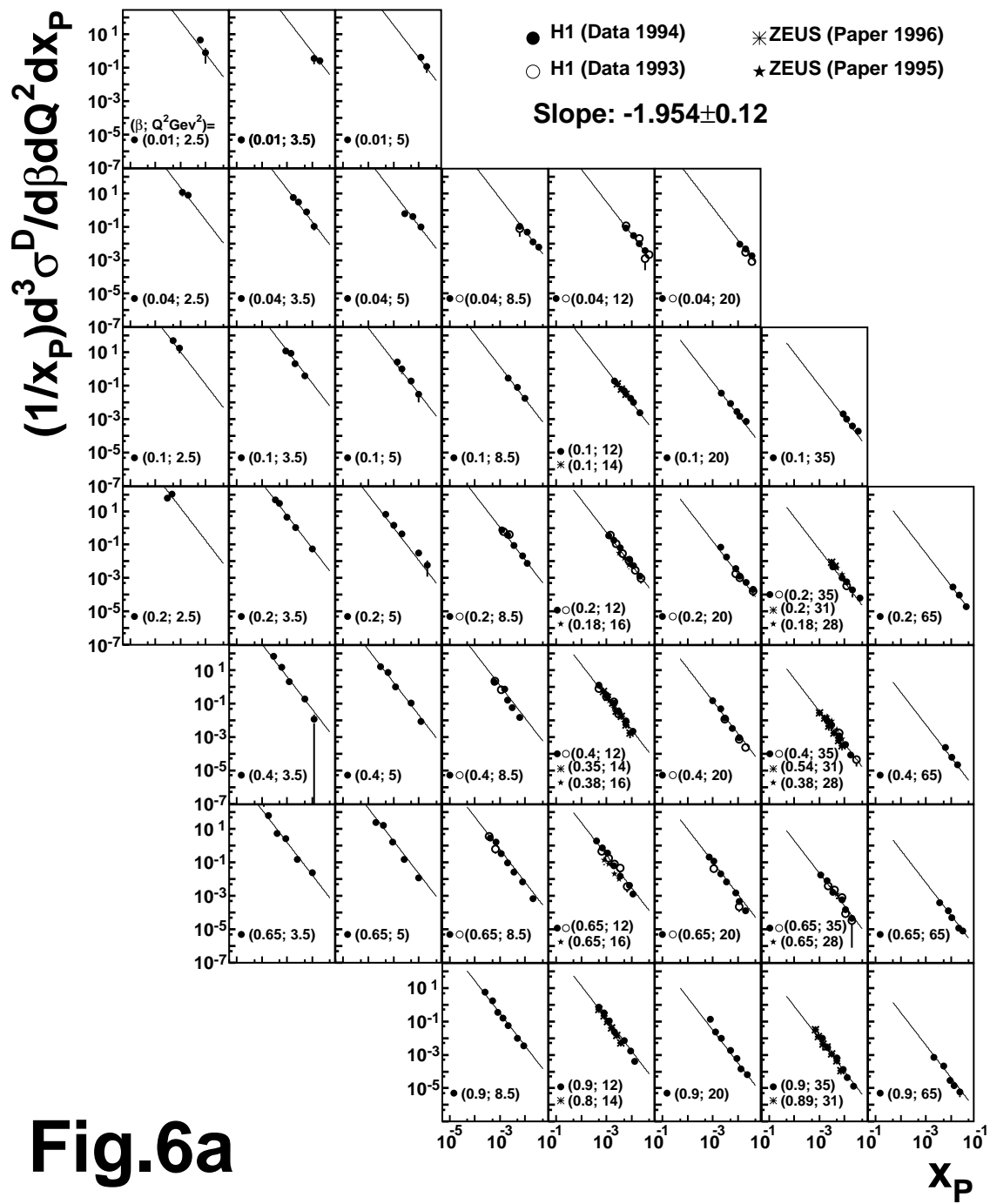
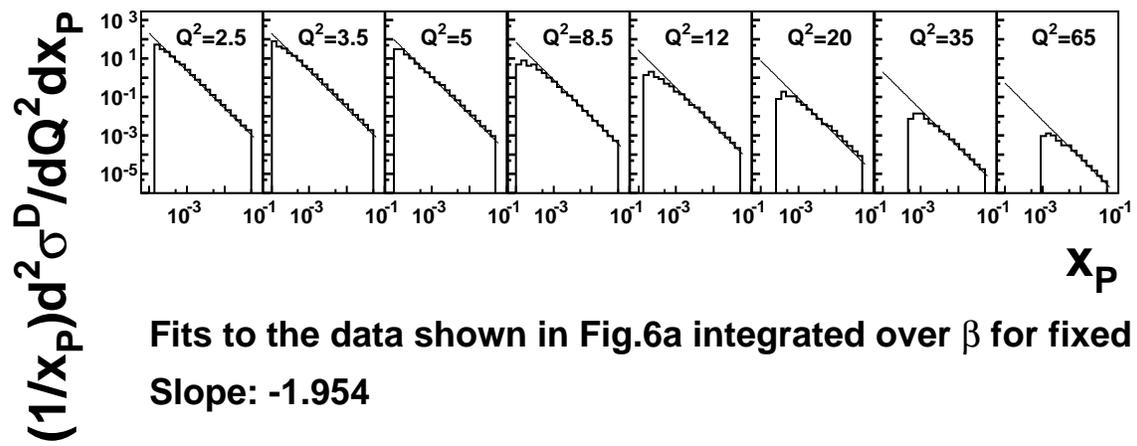
\end{document}